\documentclass[prb,twocolumn,tightenlines,floatfix,showpacs]{revtex4}
\usepackage{graphicx}

\newcommand{\beq}{\begin{equation}}
\newcommand{\eeq}{\end{equation}}
\newcommand{\bea}{\begin{eqnarray}}
\newcommand{\eea}{\end{eqnarray}}

\begin{document}

\title{Inversion of DNA charge by a positive 
polymer via fractionalization of the polymer charge}

\author{T. T. Nguyen and B. I. Shklovskii}

\affiliation{Theoretical Physics Institute, University of Minnesota, 116
  Church St. Southeast, Minneapolis, Minnesota 55455}

\begin{abstract}
Charge inversion of a DNA double helix by an oppositely 
charged flexible polyelectrolyte (PE) is widely used for gene delivery. 
It is considered here in terms of discrete 
charges of DNA. We concentrate on the worst scenario case when 
in the neutral state of the DNA-PE complex,
each of the DNA charges is
locally compensated by a PE charge and show that 
charge inversion exists even in this case. 
When an additional PE molecule is adsorbed
by DNA, its charge gets fractionalized into monomer charges of 
defects (tails and arches) on the background of the perfectly 
neutralized DNA.
These charges spread all over the DNA 
eliminating the self-energy of PE. 
Fractionalization leads 
to a substantial charge inversion of DNA. 
We show that fractionalization 
mechanism charge inversion works also for
nonlinear polymers such as dendrimers.
Remarkably, such fractionalization happens for
adsorption of both PE or dendrimers
on a two-dimensional charged lattice, as well.
Relation of fractionalization 
to other mechanisms of charge inversion is discussed.


\end{abstract}

\maketitle


\section{Introduction}

Inversion of the negative charge 
of a DNA double helix by its complexation with
a positive polyelectrolyte (PE) is 
used for the gene delivery. 
The positive charge of DNA-PE
complex facilitates DNA contact with a typically
negative cell membrane making penetration into the cell 
hundreds times more likely\cite{Kabanovs}. 
Charge inversion of DNA-PE
complexes was confirmed recently by 
electrophoresis\cite{Kabanov}.
If, at a given concentration of long DNA helices, 
the concentration of shorter PE molecules increases,
at some critical point the  
electrophoretic mobility of a DNA-PE complex 
changes sign from negative to positive.
Intuitively, one can think that when a 
PE completely neutralizes a DNA double helix
new molecules of PE stop adsorbing on DNA.
Indeed, the Poisson-Boltzmann approximation
for description of screening of a DNA helix 
by any counterions including 
PE does not lead to charge inversion. 
The counterintuitive 
phenomenon of charge inversion of a macroion by 
oppositely charged PE
has attracted significant 
attention\cite{Linse,Pincus,Bruinsma,Joanny,Joanny1,Joanny2,Shklovskii,Nguyen,Nguyen2,Nguyen3,Rubinstein,RMP}. 
It can be explained 
if one takes into account that the surface potential
of an already neutralized DNA is locally affected by 
a new approaching PE molecule, or in other words, taking into
account correlations between PE molecules\cite{Shklovskii,RMP}. 
Due to repulsive interaction between PE molecules
a new PE molecule pushes aside PE molecules which are
already adsorbed on DNA surface 
and creates on the surface 
an oppositely charged
image of itself. The image attracts the new PE 
molecule leading to charge inversion.
This phenomenon 
is similar to attraction of a charge 
to a neutral metal.

For quantitative consideration,
charges of DNA are often assumed to be smeared and 
to form uniformly charged cylinder. 
This approach seems to be justified when density 
of charge of PE is larger than density of charge at the DNA surface
so that most of DNA surface is empty.
However, it is clearly far from satisfactory when these densities 
are almost equal and 
PE charges strongly compete for charges of DNA (see figures below).
Approximation of uniform charge also 
ignores interference between 
chemical structure of DNA surface and of PE.
Therefore, generally speaking, 
it is not even clear whether charge inversion
exists in the case of discrete charges or it 
is just an artifact of the assumption 
of uniformly smeared charge. 
In this paper, we consider effects of discreteness of $-e$ charges 
of DNA. We show that in this case
charge inversion exists as well. It can be explained 
as a result of the ``fractionalization" of charge of PE molecules.
Such explanation turns out to be even simpler 
and more visual than for the model of smeared charges
of DNA.

Negative elementary
charges of DNA phosphates are situated along the two 
spirals at the exterior of both helices.
When unfolded, each spiral is an one-dimensional 
lattice of such charges, with the lattice constant $a$=6.7\AA.
Let us consider a toy model of a PE as a freely jointed chain of
$Z$ small $+e$ monomers.  
To maximize the role of
discreteness of DNA charge we begin from the assumption that 
the PE bond length $b$ is exactly equal to
the distance $a$ between negative charges of a spiral.
We call such PE ``matching". 
We also assume that minimal distance, $d$, between 
a PE charge and a charge of DNA is smaller than $a$.
Then  PE molecules can attach to a DNA charge spiral
in such a way that 
every charge of the spiral is locally compensated by a PE charge and, 
therefore, DNA is completely neutralized. 

\begin{figure}
\resizebox{8cm}{!}{\includegraphics{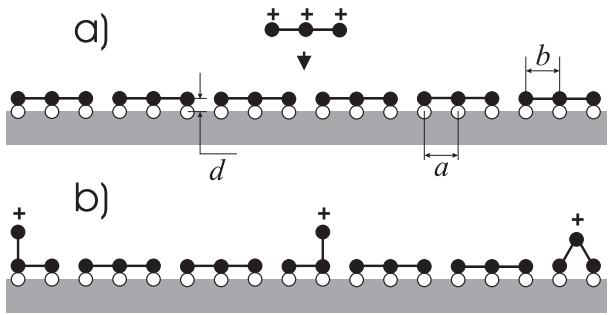}}
\caption{The origin of charge fractionalization. 
a) One of the spirals of negative charges of DNA 
(white spheres) is completely 
neutralized by positive PE molecules with $Z=3$
Their charges are shown by black spheres, neutral atoms are not shown. 
A new PE molecule is approaching DNA. 
b) The new PE molecule is "digested", 
its charge is split in $+e$ charges of two
tails and an arch (center). }
\label{fig:model}
\end{figure}

The case of a very short polymer (oligomer)
with $Z =3$ is shown in Fig.  \ref{fig:model}a 
as a simplest illustration.
The neutralization by a matching PE is so perfect that
it is difficult to imagine how another PE molecule 
can be attached to DNA. Thus, it seems
 to be impossible to overcharge DNA.
In this paper,we show that even in this  
worst possible for charge inversion scenario, there is a
mechanism which brings an additional PE to the
neutralized DNA and leads to charge inversion.
We call this mechanism fractionalization and 
Fig.  \ref{fig:model} shows how this mechanism works
for the case of $Z =3$ 
When a new PE comes to the DNA double 
helix which is already neutralized
by PE, it creates a place for itself or, in other words,
the oppositely charged image
in the following way.    

In each of $Z$ already-adsorbed PE molecules 
one PE monomer
detaches from DNA 
surface. This leads to 
formation of positive defects 
(tails and arches) and $Z$ negative 
vacancies on DNA.
All $Z$ vacancies can join together and 
form a large vacancy of a length $Z$ by
shifting of adsorbed PE molecules along DNA. A new 
PE molecule is accommodated in this vacancy.
As a result of consumption of this molecule 
$Z$ defects with charge $+e$ each appear on top 
of the  completely neutralized spiral 
(see Fig.  \ref{fig:model}b).

This effectively looks as
cutting of the new PE molecule into 
$Z$ individual monomers and spreading
them out along the spiral. In other 
words, charge inversion of DNA happens by
fractionalization of the PE molecule charge.
Of course, none of the chemical bonds 
is really cut, and this phenomenon is solely due 
to the correlated distribution of PE molecules, which avoid each other 
at the DNA spiral. In this sense,
fractionalization we are talking about is 
similar to what happens in fractional quantum Hall
effect\cite{Prange} or in the polyacetylene\cite{Braz},
where many-electron correlations result in the
fractionalization of the electron charge.

Fractionalization is driven by elimination 
of the self-energy of free PE molecules.
By the self-energy we mean the energy of
repulsive interactions of $Z$
positive charges of the PE molecule
in extended conformation which it has in the solution.
In the fractionalized state, charges of monomers 
are very far from each other and practically do not
interact, so that the positive PE
self-energy is eliminated and, therefore, gained. 

In the next section we calculate fractionalization induced 
charge inversion by a matching flexible PE. 
In Sec. III we discuss what happens when PE  does not match
the DNA spiral of charges so that 
linear densities of charge are different. In Sec. IV we generalize 
these ideas to adsorption on 
two-dimensional lattices of discrete macroion charge.
We show that in the case of matching flexible PE 
fractionalization works perfectly even in two-dimensions.
This is interesting because many other physical examples
of charge fractionalization do not work beyond one dimension.
Furthermore, this is the first classical 
example of real two-dimensional fractionalization.
In Sec. V we generalize our theory to 
flexible polymers, which do not have linear structure.
We concentrate there on charge inversion of DNA by dendrimers
and show that in this case fractionalization ideas lead
to charge inversion, too. In Sec VI, we discuss additional
mechanism of charge inversion related to the fact that
DNA charges can be accessible from two opposite sides.
We conclude in Sec. VII. A short version of this
paper is published elsewhere\cite{shortFrac}.


\section{Fractionalization induced charge inversion:
a matching polyelectrolyte}

\label{sec:marching1D}

Let us now calculate 
the linear density of the 
net charge of DNA, $\eta^*$, using the 
fractionalization mechanism. 
The chemical potential of the PE absorbed
at the spiral is
\beq
\mu_s=Zk_BT\ln(\eta^*/\eta_0) +Ze \psi(0)~.
\label{eq:muabsorbed}
\eeq
The first term in the right hand side
of Eq. (\ref{eq:muabsorbed}) is
the chemical potential of
the one-dimensional gas of defects
($-\eta_0 \simeq 0.6e/$\AA\ is the bare charge density of DNA).
We used expression for the chemical potential of an
ideal gas because the Coulomb
interaction energy between defects at the a distance of a
few $a$ is much smaller than $k_BT$ ($a \simeq l_B$, 
where $l_B = e^2/Dk_BT \simeq 7$\AA\ is the Bjerrum length.) 
The second term in the right hand 
side of Eq. (\ref{eq:muabsorbed}) 
is the repulsion energy of the new PE
from the inverted charge of the DNA. 
In this term, $\psi(0)$ is
the averaged surface potential of the DNA helix. 
We assume in this paper that the net charge 
of DNA is screened by a monovalent salt
at the screening length $r_s$, which is much larger than $a$.
Then $\psi(0)$ can be calculated as
the surface potential
of a cylinder with radius of DNA helix $R$ 
and linear density of charge $\eta^*$
\beq
\psi(0)\simeq \frac{2\eta^*}{D}\ln\frac{r_s+R}{R}~.
\eeq
To find $\eta^*$ in the equilibrium state, 
one has to equate the chemical potential of adsorbed
PE molecules with that of a free PE in the solution. 
The later one can be
calculated as following. Due to the
repulsive Coulomb interaction between monomers,
a free PE in the solution has an extended shape to minimize its energy.
Therefore, the chemical potential
of a free PE in solution can be written as the self-energy of 
a rigid rod with the length $Za$ and the linear charge density $e/a$
plus the standard ideal gas contribution $k_BT\ln(Nv_0)$ ($N$ is the
number concentration of free PE in solution and
$v_0$ is the volume of a PE molecule):
\beq
\mu_0= (Ze^2/Da)\ln({\cal L}/a)+k_BT\ln(Nv_0)~,
\label{eq:mufree}
\eeq
where ${\cal L}=\min(r_s,Za)$ and 
$D$ is the dielectric constant of water.

Equating the chemical potentials of
Eqs. (\ref{eq:mufree}) and  (\ref{eq:muabsorbed}),
one has
\beq
\psi(0) =\frac{e}{Da}\ln\frac{\cal L}{a}+ 
\frac{k_BT}{e}\ln\frac{\eta_0}{\eta^*}+
\frac{k_BT}{Ze}\ln(Nv_0)~.
\label{eq:charging}
\eeq
In Eq. (\ref{eq:charging}) one can interpret the right hand 
side as a ``correlation" voltage 
that (over)charges the DNA to the potential $\psi(0)$.
Complete analysis of Eq. (\ref{eq:charging}) is given in the
Appendix. 
It shows that with growing $N$ the net charge of DNA $\eta^*$
experiences a first order transition from
negative to positive values.
Here we concentrate only at large enough $N$,
where $\eta^*$ is positive.

Let us make two simplifying approximations.
Firstly, we assume that the concentration $N$
of PE in the solution is large enough
so that PE translational entropy term (the last
term in Eq. (\ref{eq:charging})) can
be neglected. In other words,
we calculate the maximum possible charge inversion.
This limit is reached when $N \gg N_0$, where 
\beq
N_0=v_0^{-1}\exp(-Zl_B\ln({\cal L}/a)/a)
\label{eq:N_0}
\eeq
is an exponentially small characteristic concentration. For a long PE
$N_0$ is so small that one does not need alarge $N$ to get to this limit.

Secondly, as a good approximation, one can now neglect
second term of the right side of Eq. (\ref{eq:charging}), 
which is responsible 
for the entropy of defects on DNA. This easily
leads to a solution for the net charge density
\beq
\eta^*\simeq\frac{e}{2a}\frac{\ln({\cal L}/a)}{\ln[(r_s+R)/R]}~.
\label{eq:eta*}
\eeq
Now one can check that this solution
is consistent with the assumption that the entropic term can be 
neglected by substituting it back into
Eq. (\ref{eq:charging}). 

Equation (\ref{eq:eta*}) shows that $\eta^*$ is positive 
indicating that the bare DNA charge is inverted. 
Knowing $\eta^*$ and using $|\eta_0| = 0.6e/$\AA$ \simeq 3.9 e/a$
the charge inversion ratio can be calculated
\beq
\left|\frac{\eta^*}{\eta_0}\right| = 
0.13 \frac{\ln({\cal L}/a)}{\ln[(r_s+R)/R]} ~.
\label{eq:result}
\eeq
For DNA $R = 10$\AA\ and $a = 6.7$\AA, so that at $r_s \geq 10$
\AA\ the ratio of logarithms 
can be only slightly larger than unity. Thus, the
charge inversion ratio created by fractionalization is limited by 
20\%.  Up to such point we indeed can neglect
Coulomb interactions between defects in the chemical potential
of the gas of defects (the first term in the right hand side
of Eq. (\ref{eq:muabsorbed})).

Remarkably, the extremely crude bead-and-stick
model of PE discussed above can give reliable and
universal predictions.
The calculation described above is not sensitive
to many microscopic details and chemically-specific
effects on atomic scale. One could
worry about behavior of dielectric constant of water
at small distances, destruction of water
solvation shells, other interactions (van der Waals, hydrogen bonds,
etc.)
All of them are not
important because they all modify energy of interaction
of PE with DNA which {\em does not enter} in the above calculation.
This energy is identical for configurations
on Fig.  \ref{fig:model}a and Fig.  \ref{fig:model}b.
The only difference between these configurations
is the self-energy of a free PE molecule,
which does not depend on any details of the PE-DNA interaction.
Only this self-energy drives charge inversion.

One could also ask about the role of the finite
flexibility of PE for the tails. As we all know,
freely jointed chain model of polycation is useful on
length scales of several nanometers, but is not literally valid even
on length scales of 6-7 \AA. We want to
emphasize that we do not need ideal flexibility of
tails, which lets them to be perpendicular to DNA cylinder surface. The
only requirements for flexibility of tails assumed in our calculation
is that the tail can be raised in such a way
that its end monomer
avoids the end monomer the neighboring PE molecule.
This requirements is fulfilled in many cases,
for example, for the spermine\cite{Spermine} ($Z=4$).
(One should
take into account that the neighboring PE charged monomers
are usually connected by a chain of several neutral monomers).

Small arches shown on Fig.  \ref{fig:model}b,
however, are more sensitive to flexibility than tails.
If the persistent length of PE, $l$ is larger 
than the distance between charges, $a$,
loops (arches) have a typical length $l$. In a long PE where 
arches dominate this leads to replacement of $\ln({\cal L}/a)$
by $\ln({\cal L}/l)$ in Eq. (\ref{eq:result}) and therefore to 
a somewhat weaker charge inversion. 

In the mostly theoretical case of a short and extremely rigid 
PE when even tails can not bend at all, so that a PE charge of a 
neutralized DNA is totally incompressible,
both fractionalization and charge inversion 
disappear. This is  similar to what happens when $Z$-ions 
are hard spheres and one layer of them exactly compensates the
uniformly charged background\cite{Nguyen3}. Charge inversion disappeares
in these cases, because there are no internal degrees of 
freedom of molecules to make the system compressible.

Until now we talked about one-dimensional periodic
chain of negative charges. If we recall that in DNA this chain actually 
is a spiral we face another requirement for 
the flexibility of a long PE.
A PE molecule should be flexible enough to follow DNA spiral.
Most of PE can do that, for example spermine does\cite{Spermine}.
On the other hand, extremely rigid long PE can
not follow a spiral of charge and, therefore, screens DNA 
as an uniformly charged cylinder, namely
PE rods in this case arrange themselves at its surface 
collinearly with the cylinder axis and each other.

Concluding this section, we would like
to say that the discreteness of charges does not prevent 
charge inversion even in the worst case of perfect matching.


\section{Polyelectrolyte with non-matching density of charge}

\begin{figure}
\resizebox{8cm}{!}{\includegraphics{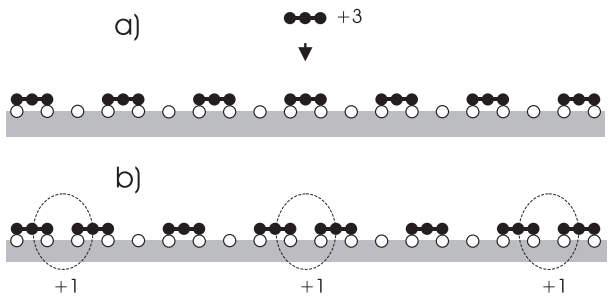}}
\caption{The origin of charge fractionalization 
for a PE with linear charge density twice larger than for 
a DNA spiral. a) A Wigner-crystal-like ground state 
of a periodic chain of negative charges neutralized by PE
molecules with $Z=3$ and $b = a/2$
(their charges are shown by black spheres). 
A new PE molecule is approaching DNA. 
b) The new PE molecule is "digested" 
by DNA. Its charge is split in $+e$ charges of $Z$ grain boundaries.}
\label{fig:twice}
\end{figure}

How does fractionalization 
work when distance between charges of PE,
$b$, is not equal 
to the distance between charges of an unfolded DNA spiral, $a$?
Consider, for example, commensurate
PE with $b = a/2$, which has linear density of charge twice 
larger than a DNA spiral. 
In this case, PE molecules due to Coulomb repulsion form an analog of
Wigner crystal where PE molecules alternate with vacant places
(see Fig.  \ref{fig:twice}a).
Even if the PE is absolutely rigid
 a new PE molecule creates $Z$ distant grain boundaries 
(domain walls), where
one vacancy is missing (see Fig.  \ref{fig:twice}b). 
The charge of each grain boundary 
is $+e$, so that charge of the new PE molecule is fractionalized.
and a part of the self-energy of PE is eliminated in the way 
similar to what happens in the case of matching PE. 

\begin{figure}
\resizebox{8cm}{!}{\includegraphics{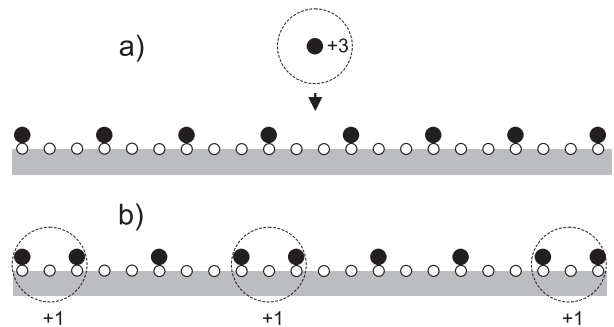}}
\caption{The origin of charge fractionalization 
for multivalent counterions. a) A Wigner-crystal-like ground state 
of a periodic chain of negative charges (white spheres) neutralized by 
multivalent counterions with $Z=3$ (larger black spheres). 
b) The new $Z$-ion is "digested" 
by DNA. Its charge is split in $+e$ charges of $Z$ grain boundaries.
Broken circles are explained in the text}
\label{fig:Lantanum}
\end{figure}

Fractionalization continues to work
when the linear charge density of multivalent counterion ($Z$-ion)
is even larger. We can imagine such limit, when replacing PE with a
metallic multivalent ion (for example, $La^{+3}$), 
which touches only one negative charge 
of DNA. Then we arrive at a ground state of neutralized DNA
which resembles Wigner crystal even closer (see Fig.  \ref{fig:Lantanum}a).
Fractionalization of a new charge into $Z$ 
monovalent charges of grain boundaries
(see Fig.  \ref{fig:Lantanum}b) decreases self-energy
and drives charge inversion.
In contrary to obvious elimination of 
interaction between monomers in the case of Fig.  \ref{fig:model}.
it is more difficult to see how self-energy is eliminated 
in Figs.  \ref{fig:twice} and \ref{fig:Lantanum},
For example, to get an idea how this happens 
in the case of Fig.  \ref{fig:Lantanum} it helps to 
draw a sphere with radius a bit larger than $a$ both around 
the new free $Z$-ion on Fig.  \ref{fig:Lantanum}a
and around the center of each of $Z$ domain walls on 
Fig.  \ref{fig:Lantanum}b.
They are shown by broken circles. Let us consider now what 
happened to the energy of the electric field 
of the new $Z$-ion concentrated in the external space of these spheres. 
Due to fractionalization of $Z$-ion 
the energy of the charge $Z$ is clearly 
replaced by the smaller sum of $Z$ energies of monovalent ions.
This illustrates what we mean
talking about elimination of the self-energy in this case.

In Fig.  \ref{fig:Lantanum} we already arrived at a model of 
charge density wave and fractionalization in polyacetilene
and a very crude picture for the fractional Hall effect at filling factor 1/3. 
In the latter case, empty circles mean discrete Landau states
and an electron  charge $-e$ is split in 3 charges $-e/3$.
  
Fig. \ref{fig:Lantanum} also resembles what happens in the case 
of adsorption of $Z$-ions on the line or surface with uniform 
distribution of background charge \cite{Nguyen,Shklovskii}.
In that case, the charge of a new $Z$-ion is 
smeared along the background due 
to small elastic deformations of
Wigner-crystal-like strongly correlated liquid.
In other words, $Z$-ion is fractionalized
into infinitesimally small portions.
One can visualize the transition to the case of uniform surface 
charge imagining that both elementary charge of our 
lattice and lattice constant 
$a$ vanish, while charge density of DNA
and charge of $Z$-ion are kept constant. 

Let us return to adsorption of PE with a finite
linear charge density on DNA 
and discuss more complicated situations,
when $b<a$, but $b$ and $a$ are incommensurable.
Even in this case ground state of 
a neutralized DNA is a crystal. If an additional PE molecule 
is adsorbed it is still fractionalized 
to $Z$ grain boundaries with charge $+e$.
The only difference from commensurable case shown on 
Fig.  \ref{fig:twice} is that grain boundary can 
include several PE molecules.


\section{Fractionalization of polyelectrolyte
charge in two-dimensions}

\begin{figure}
\resizebox{8cm}{!}{\includegraphics{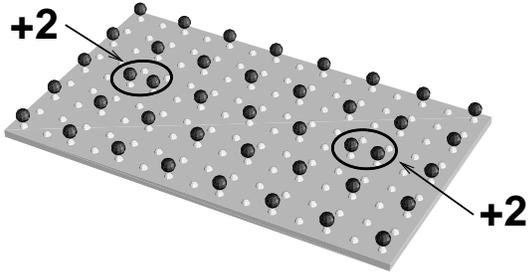}}
\caption{The origin of a shear deformation string between
two $+2e$ defects,
which appeared as a results of adsorption of an
additional $Z$-ion  with $Z=4$ in a two-dimensional
generalization of the model of Fig.  \ref{fig:Lantanum}.}
\label{fig:str2D}
\end{figure}

It is well known that 
fractionalization of charge of $Z$-ion
into free (to move to infinity) grain boundaries 
shown for example in Fig. \ref{fig:Lantanum}
can not be generalized to a two-dimensional
case. Let us imagine a two-dimensional analog 
of the problem of Fig. \ref{fig:Lantanum} 
using a square lattice of monovalent negative charges 
which is neutralized by $Z$-ions with charge $Z=4$ 
forming a square lattice with period $2a$.
If we bring another $Z$-ion and try to split 
it into two point like grain boundaries with charge $+2e$
along the main axes of the square lattice,
we realize that 
the lattice of $Z$-ions 
looses energy everywhere between them because
of the shear deformation created (See Fig.  \ref{fig:str2D}).
This is equivalent to a string between charges $+2e$
with energy proportional to length.
Thus, charges $+2e$ of the defects are not permitted
to move very far away from each other. 
In other words, they are confined in a 
finite domain. 

Nevertheless, in the first approximation, 
charge inversion can be still calculated as if
products of the $Z$-ion fractionalization
were free to move to infinity. 
Indeed, one can estimate the defect confinement 
size and find that it is 
much larger than $a\sqrt{Z}$ (the average distance
between $Z$-ions on the surface), because the energy 
of abovementioned strings is 
proportional to
the shear modulus of the 
Coulomb lattice of $Z$-ions on the 
negative background lattice
which is known to be numerically small.
Therefore, most of the self-energy 
of $Z$-ion concentrated in the electric field 
at radius larger than $a\sqrt{Z}$ is eliminated in spite of 
defects confinement.

\begin{figure}
\resizebox{8cm}{!}{\includegraphics{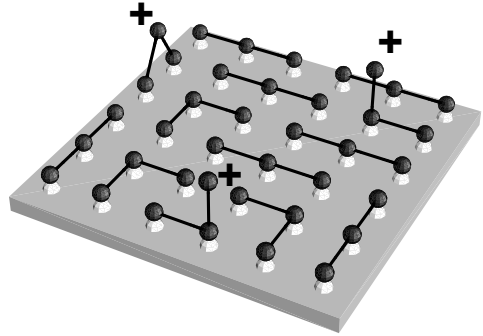}}
\caption{Fractionalization of an additional PE 
molecule with $Z=3$ into three positive defects 
at a
two-dimensional lattice of negative charges (white spheres)
neutralized by already adsorbed PE.
Positive charges of PE molecules are shown by black spheres.}
\label{fig:2DFrac}
\end{figure}

Remarkably, 
for a reasonably flexible matching PE 
fractionalization into free tails and arches
is not a strictly one-dimensional phenomenon.
It is easy to see that the same 
mechanism applies equally well to
a two-dimensional square lattice of discrete 
negative charges with the lattice constant, $a$, equal
to the PE bond length $b$. 
Indeed, one can see in Fig. \ref{fig:2DFrac} that
all previous arguments about the role of tails and arches
can be carried over to this case.
There are no strings between tails and
arches in this case.
This is a remarkable consequence
of the involvement of additional
degrees of freedom related to the third dimension.
We do not know any other classical example
of charge fractionalization in a really
two-dimensional system.

There are, however, small modifications of the 
analytic formulae for charge inversion.
Defects with $+e$ charges 
form now a two-dimensional gas with
concentration
$\sigma^*/e$, where $\sigma^*$ is the net
positive surface charge density playing the
roles of $\eta^*$. The chemical 
potential of this gas is $k_BT\ln(a^2\sigma^*/e)$.
The surface potential is 
$\psi(0)=2\pi\sigma^* r_s/D$.
The balance of the chemical potential of PE molecules
adsorbed at the surface with that of a free PE 
in the solution reads
\beq
\frac{2\pi\sigma^*r_s}{D}=\frac{e}{Da}\ln\frac{\cal L}{a}+
\frac{k_BT}{e}\ln\frac{e}{a^2\sigma^*}+
\frac{k_BT}{Ze}\ln(Nv_0)~.
\label{eq:psi02d}
\eeq
Again, assuming that the PE concentration $N$ is large
(or calculating the maximum possible charge inversion)
the solution to Eq. (\ref{eq:psi02d}), 
for $a \simeq l_B$, within a numerical factor, is
\beq
\sigma^* \simeq (e/ar_s)/\ln(r_s/a)~.
\eeq
One can see that, for $a \simeq l_B$, in the free
energy gained by fractionalization of 
the PE molecule charge, the entropy
contribution is comparable to the
self-energy, in contrary with
the one-dimensional case, where the entropic term can be neglected.
This is due to a higher number of degrees of freedom 
which a 
two-dimensional surface provides to the gas of defects.
If $r_s \gg a$, the charge 
inversion ratio for the two-dimensional 
case is smaller than for DNA:
\beq
\left|\frac{\sigma^*}{e/a^2}\right|
=\frac{a}{r_s}\ln\frac{r_s}{a} ~.
\eeq
An important role of elimination of the self-energy 
for adsorption of a flexible 
PE on an oppositely uniformly charged surface 
can be traced in Refs. \onlinecite{Pincus}, 
\onlinecite{Joanny}, \onlinecite{Rubinstein}.


\section{Charge inversion of DNA by dendrimers and fractionalization}

\begin{figure}
\resizebox{8cm}{!}{\includegraphics{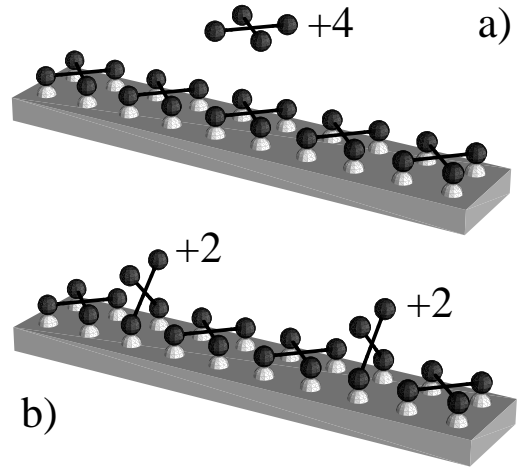}}
\caption{The origin of charge fractionalization in 
dendrimer 
adsorption. a) Two linear chains of negative charges of DNA 
(white spheres) which are obtained 
unfolding DNA spirals of charges. They are completely 
neutralized by positive dendrimer molecules with $Z=4$.
Dendrimer are schematically shown by plane
crosses with +e charges (black spheres) at the ends. 
A new dendrimer molecule is approaching DNA. 
b) The new dendrimer molecule is "digested" 
by DNA. Its charge is split in $+2e$ charges of the tail
doublets.}
\label{fig:dend}
\end{figure}

Until now we considered adsorption 
of linear charged molecules (PE)
both on one- and two-dimensional lattices of the
background charge.
It is interesting to note that 
the fractionalization mechanism works 
for molecules of other shapes, too.
Let us, for example, consider dendrimers 
(star-like branching molecules 
with  a large number of monovalent positive charges 
on their periphery), which were
also shown to invert the charge of DNA
\cite{KabanovD}. Dendrimers
with charges $Z$=4, 8 can easily compensate
a compact group of nearest $Z$ charges of both DNA helices.
If a DNA double helix is totally covered and neutralized
by such dendrimers (see the schematic Fig.  \ref{fig:dend}a for $Z$=4)
an additional dendrimer can still be adsorbed on 
DNA. This happens because two charges $+e$ 
of two distant
already adsorbed 
dendrimers can be raised above the DNA surface 
when a new 
dendrimer molecule is adsorbed on DNA 
(see Fig.  \ref{fig:dend}b).
As in the case of linear molecules, this 
fractionalization of the dendrimer with charge $+4e$ 
into two charges $+2e$
leads to the gain of its self-energy and 
to charge inversion. 

Again we see that all these phenomena 
became possible only due to the additional 
of freedom of PE molecules, which in this case is rotational. 
(Fractionalization 
into charges $+e$ in this case can leads to a larger energy because
all adsorbed dendrimers
between two dendrimers raising one tail
should be deformed leading to a string with energy 
proportional to the length between them.) 

\begin{figure}
\resizebox{8cm}{!}{\includegraphics{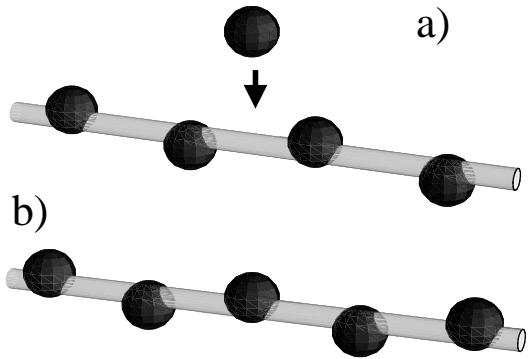}}
\caption{The origin of charge inversion in 
adsorption of high generation dendrimers. a) DNA double helix 
(gray) neutralized by a
Wigner-crystal-like liquid of a high generation 
dendrimers (dark spheres). A new dendrimer molecule is approaching DNA. 
b) The new dendrimer molecule is integrated into 
Wigner-crystal-like liquid while neighboring already adsorbed dendrimers
slide away from it and smear its charge over the helix.}
\label{fig:dendhigh}
\end{figure}

If we deal with higher generations of 
dendrimers which have very large charges such as $32e$ or $64e$,
we arrive at a different Wigner-crystal-like picture
(see Fig.  \ref{fig:dendhigh}).
Because of the three-dimensional structure of 
their chemical bonds these molecules can 
not expand enough so that 
each charge of them reaches an opposite charge of DNA
and compensates it. In other words,
when projected to a DNA double helix, these high generation dendrimers
have much larger linear density
of charge than the double helix itself. Thus, 
large segments of the helix 
between adsorbed dendrimers remain negatively charged, 
and form a Wigner-Seitz cells around each dendrimer.
This is how with growing charge 
of dendrimers the fractionalization mechanism is replaced by the
mechanism of Wigner-crystal-like correlations. 
Qualitative difference between DNA complexes with dendrimers of low 
and high generations  
has been clearly demonstrated experimentally\cite{KabanovD}. 
Because large fraction of DNA charges is not neutralized by dendrimers
the high generation complexes 
are more sensitive to the salt concentration.


\section{Sharing of DNA charges  
as a mechanism of charge inversion}

\begin{figure}
\resizebox{8cm}{!}{\includegraphics{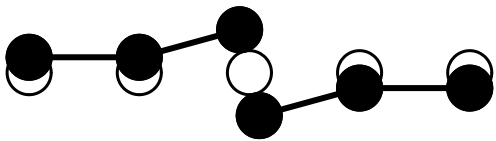}}
    \caption{A view from the top on an unfolded spiral of negative charges
of DNA (white spheres) and two PE molecules (black).
Two positive end monomers share a negative charge of DNA in the perfect 
opposition.}
    \label{fig:twoplus}
\end{figure}

Let us return to complexation of a DNA 
double helix with PE molecules with the matching
bond length, $b = a$, and discuss another possible
mechanism of charge inversion, 
which is also related to the discreteness
of DNA charge and further increases 
the positive charge of DNA-PE complex. 
Let us consider a monomer tail of PE on 
Fig.  \ref{fig:model}b and explore 
whether some energy 
can be gained if the positive charge of this monomer 
moves down to the plane of DNA charges,
approaches already neutralized negative charge of the DNA and  
shares it with the end monomer of the neighboring PE molecule 
in a way shown in Fig.  \ref{fig:twoplus}.
If these two end monomers may sit on exactly opposite sides of 
the negative charge of DNA, the 
additional energy $e^2/2d$ can be gained,
where $d$ is the distance of the 
closest approach of a PE monomer and a DNA charge.
At a sufficiently small $d$ this energy
can be even larger 
than the gain per tail from elimination of the self-energy.
In a DNA double helix, all the
negative charges indeed are on the ridge above 
neighboring neutral atoms. Two sufficiently small 
monomers may fit into 
the large and small groves on both sides of the ridge.
On the other hand, if because of sterical limitations
they can not be in the perfect opposition the 
energy gain is smaller. If both end monomers PE have the same size 
as the negative charge of DNA 
the additional energy of sharing vanishes when 
all three spheres touch each other forming 
equilateral triangle. This still 
leaves room for sharing effect, while say one 
monomer perfectly fits in large grove but the second one 
only partially fits in the small grove.
%


\section{Conclusion}

The main result of this paper is that discreteness of 
surface charges of a macroion, for example, double helix 
DNA does not 
prevent its overcharging by oppositely charged 
PE and other $Z$-ions. For a flexible PE
even in the worst scenario of matching PE
and DNA geometries, when in the neutral state all charges of DNA 
are perfectly neutralized by PE, charge inversion 
happens due to fractionalization of PE charge into 
$+e$ charges of defects. This is extremely 
transparent mechanism as illustrated on 
Fig.  \ref{fig:model}. 
It is clearly related to internal  
degrees of freedom of a flexible PE.
In the non-matching cases, the
mechanism of charge inversion for discrete surface charges 
looks more similar to the one
previously discussed in a model of uniformly charged
macroion surface, but still is accompanied by fractionalization.

In conclusion, we emphasize that in any case
charge inversion happens due to the fact that a new 
PE molecule rearranges already
adsorbed PE in such away that its 
image or correlation hole strongly 
attracts this new PE molecule. 
This physics can not be described 
by the Poisson-Boltzmann
theory because this theory uses
the mean-field potential which 
does not depend on the position of a new PE molecule.

\begin{acknowledgments}

We are grateful to A. V. Kabanov for 
the question which initiated this work, 
and A.\ Yu.\ Grosberg, V. A. Kabanov and  
P.\ Pincus for useful discussions of results.
This work was supported 
by the NSF grant No. DMR-9985785.
T.T.N. is also supported by the Doctoral Dissertation 
Fellowship of the University of Minnesota.
\end{acknowledgments}


\appendix
\section{Charge of DNA as a function of polyelectrolyte concentration}

In Sec. \ref{sec:marching1D}, we assumed the bulk PE concentration, $N$,
is very large so that the translation 
entropy cost of condensing them on DNA
can be neglected. Equation (\ref{eq:eta*}), thus, gives
the upper limit for the DNA net inverted charge. On the other hand,
at very small $N$, the entropy cost cannot be neglected and
leads to the undercharging of DNA.
In this appendix, we would like
to calculate $\eta^*(N)$ explicitly and
show that in the matching case
DNA molecules change their sign with increasing $N$ 
by a first order phase transition. 

When DNA is undercharged, $\eta^*<0$, instead of a gas of 
raised monomers (tails and arches) on the DNA surface,
one has a gas of vacancies. These are the DNA charges which are
not covered by any PE monomers. At low concentration 
(small undercharging), these vacancies
practically do not interact and their chemical
potential can be approximated by that of an ideal gas
at the same concentration $k_BT\ln(\eta_0/|\eta^*|)$.
Thus, Eq. (\ref{eq:charging}) needs only a small modification
to properly describe both the over- and under-charged DNA:
\beq
\frac{2\eta^*}{e/l_B}\ln\frac{r_s+R}{R}-
\ln\frac{\eta_0}{|\eta^*|}=\frac{1}{Z}\ln\frac{N}{N_0}
\label{eq:charging3}
\eeq
where the second term is the chemical potential of raised
monomers in the overcharging case and it is the chemical potential
of vacancies in the undercharging case. 
In Eq. (\ref{eq:charging3}), we have also combined two terms
of Eq. (\ref{eq:charging}) 
using the characteristic concentration $N_0$
given by Eq. (\ref{eq:N_0}).

It should be noted that,
the apparent divergence of the left side of
Eq. (\ref{eq:charging3}) at small $\eta^*$
is related to the fact that we neglected
a small concentration of intrinsic
defects (raised monomers and vacancies).
This concentration is of the order $a^{-1}\exp(-e^2/2dk_BT)$.
It exists even at $\eta^*=0$ and truncates the divergence of 
$\ln(\eta_0/|\eta^*|)$ at small $\eta^*$.
However, when $|\eta^*| \gg ea^{-1}\exp(-e^2/2dk_BT)$
one type of defects dominates over the other and
one can neglect the contribution from the minority ones.
This is what we did in Eq. (\ref{eq:charging3}).

To understand how $\eta^*$ varies with $N$, it is
very instructive to solve Eq. (\ref{eq:charging3}) graphically.
One can see the following behavior:
\begin{figure}
\resizebox{8.5cm}{!}{\includegraphics{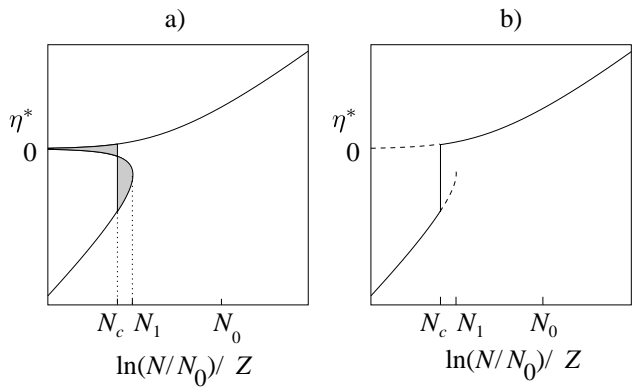}}
\caption{Behavior of $\eta^*(N)$ near the charge
inversion transition. The limit discussed in
Sec. \ref{sec:marching1D} starts only in
the upper-right corner of the figures.
a) Graphical solution to Eq. (\ref{eq:charging3}). The solid
line is the right hand side of Eq. (\ref{eq:charging3}).
When $N>N_1$, there is only one solution $\eta^*$.
When $N<N_1$, there are three solutions, $\eta^*_{1,2,3}$.
b) The DNA net charge density $\eta^*$ as a function
of $N$ (the solid line). A first order phase transition from the
undercharged to overcharged state happens at $N=N_c$.
The dashed lines correspond to the metastable values of
$\eta^*$.}
\label{fig:etaN}
\end{figure}

When $N$ is large such that $N>N_1$, where 
$\ln(N_1/N_0)/Z=-1-\ln\{2\eta_0\ln[(r_s+R)/R]/(e/l_B)\}$,
Eq. (\ref{eq:charging3}) has only one solution for $\eta^*$.
This solution is positive, indicating that the DNA helix
is overcharged.

When $N$ decreases a little bit below $N_1$, there
are three solutions $\eta^*_1 > 0 > \eta^*_2 > \eta^*_3$
of Eq. (\ref{eq:charging3}) (Fig. \ref{fig:etaN}a).
The solution $\eta^*_1$ corresponds to the stable
overcharged state. The solution $\eta^*_3$ corresponds to
the metastable undercharged state. The solution $\eta^*_2$
is unstable (it corresponds to a local
maximum in the grand potential of the system located between
two local minimums at $\eta^*_1$ and $\eta^*_3$.)

When $N$ decreases below $N_c$ where $N_c$ is defined
as the PE concentration at which the two shaded
areas in Fig. \ref{fig:etaN}a
equal each other (Maxwell rule), a first
order phase transition happens. 
(Note that
at $\exp(-e^2/2dk_BT) \ll 1$
a calculation of $N_c$ can be done with help
of Eq. (\ref{eq:charging3})
because the truncation due to intrinsic
defects produces only a small correction to one of
areas.)
The overcharged solution
$\eta^*_1$ becomes metastable while the undercharged
solution  $\eta^*_3$ becomes stable. Thus,  
the function $\eta^*(N)$ has a finite jump at $N = N_c$. This
function is plotted by the solid line in Fig. \ref{fig:etaN}b.
The metastable branches of $\eta^*(N)$ are plotted as
the dashed line.

\end{document}